\documentclass[reprint, amsmath,amssymb, aps, prb, superscriptaddress, nofootinbib, nobibnotes]{revtex4-1}

\usepackage{graphicx}
\usepackage{dcolumn}
\usepackage{bm}
\usepackage{scrextend}
\usepackage{hyperref}
\usepackage{tabularx}
\usepackage{amsmath}
\usepackage{float}
\newcommand{\one}{{\ensuremath{\pmb{k}}}}
\newcommand{\two}{{\ensuremath{\pmb{q}}}}
\usepackage{filecontents}
\usepackage{color}
\usepackage{multirow}
\usepackage[T1]{fontenc}

\begin{document}

\title{Towards temperature-induced topological phase transition in SnTe:\\ A first principles study}

\author{Jos\'e D. Querales-Flores}
\email{jose.querales@tyndall.ie}
\affiliation{Tyndall National Institute, Lee Maltings, Dyke Parade, Cork T12 R5CP, Ireland}
\author{Pablo Aguado-Puente}
\affiliation{Atomistic  Simulation  Centre,  Queen's  University  Belfast, BT7 1NN  Belfast,  Northern  Ireland,  United  Kingdom}
\author{{\DJ}or{\dj}e Dangi\ifmmode \acute{c}\else \'{c}\fi{}}
\affiliation{Tyndall National Institute, Lee Maltings, Dyke Parade, Cork T12 R5CP, Ireland}
\affiliation{Department of Physics, University College Cork, College Road, Cork T12 K8AF, Ireland}
\author{Jiang Cao}
\affiliation{Tyndall National Institute, Lee Maltings, Dyke Parade, Cork T12 R5CP, Ireland}
\affiliation{School of Electronic and Optical Engineering, Nanjing University of Science and Technology, Nanjing, 210094, China}
\author{Piotr Chudzinski}
\affiliation{Atomistic  Simulation  Centre,  Queen's  University  Belfast, BT7 1NN  Belfast,  Northern  Ireland,  United  Kingdom}
\author{Tchavdar N. Todorov}
\affiliation{Atomistic  Simulation  Centre,  Queen's  University  Belfast, BT7 1NN  Belfast,  Northern  Ireland,  United  Kingdom}
\author{Myrta Gr\"uning}
\affiliation{Atomistic  Simulation  Centre,  Queen's  University  Belfast, BT7 1NN  Belfast,  Northern  Ireland,  United  Kingdom}
\affiliation{European Theoretical Spectroscopy Facility}
\author{Stephen Fahy}
\affiliation{Tyndall National Institute, Lee Maltings, Dyke Parade, Cork T12 R5CP, Ireland}
\affiliation{Department of Physics, University College Cork, College Road, Cork T12 K8AF, Ireland}
\author{Ivana Savi\'c$^{1,}$}
\email{ivana.savic@tyndall.ie}

\date{\today} 

\begin{abstract}

The temperature renormalization of the bulk band structure of a topological crystalline insulator, SnTe, is calculated using first principles methods. We explicitly include the effect of  thermal-expansion-induced modification of electronic states and their band inversion on electron-phonon interaction. We show that the direct gap decreases with temperature, as both thermal expansion and electron-phonon interaction drive SnTe towards the phase transition to a topologically trivial phase as temperature increases. The band gap renormalization due to electron-phonon interaction exhibits a non-linear dependence on temperature as the material approaches the phase transition, while the lifetimes of the conduction band states near the band edge show a non-monotonic behavior with temperature. These effects should have important implications on bulk electronic and thermoelectric transport in SnTe and other topological insulators. 

\end{abstract}

\maketitle

\section{Introduction}\label{introduction}

SnTe~\cite{tanaka2012} and its substitutional alloys (Pb$_{1-x}$Sn$_x$Te~\cite{xu2012} and Pb$_{1-x}$Sn$_x$Se~\cite{story2012}) belong to a particular class of topological materials known as topological crystalline insulators (TCI)~\cite{Hsieh2012}. Unlike the related trivial semiconductors PbTe and GeTe, SnTe has inverted valence and conduction bands at four equivalent L points in the first Brillouin zone. This leads to an even number of Dirac-like conducting states on (001), (110) or (111) surfaces that are protected by mirror symmetry~\cite{Hsieh2012}. In contrast, $Z_2$ topological insulators have an odd number of Dirac surface states protected by time-reversal symmetry~\cite{fu2007,hasan2010,qi2011}. SnTe-based alloys can undergo a phase transition between the TCI and normal insulator (NI) phases by varying pressure~\cite{ong2017,barone2013}, temperature~\cite{story2012,wojek2014,guldner2018}, or alloy composition~\cite{xu2012,tanaka2013,story2018,guldner2018}. The topological phase transition occurs when the band gap closes and the band inversion between valence and conduction bands disappears.  The gap vanishes only at the transition temperature and the material is semiconducting on either side of the transition. Temperature-driven topological phase transitions may be especially important for devices based on these materials. Recent angle-resolved photoemission spectroscopy and magneto-optical Landau level spectroscopy measurements have shown that such a transition between TCI and NI phases can indeed be realized in Pb$_{1-x}$Sn$_x$Se~\cite{story2012,wojek2014,guldner2018}.

A potential thermally induced topological phase transition in SnTe~\cite{kim2015} may significantly affect its thermoelectric transport properties. SnTe is an emerging thermoelectric material that reaches a maximal thermoelectric figure of merit between 0.6 and 1.4 at intermediate temperatures ($600$-$1000$~K) at relatively high intrinsic hole concentrations ($\sim$ 10$^{20}$ cm$^{-3}$)~\cite{Zhang2013,Snyder2014,tan2014,tan2015,banik2015,zhao2016,alorabi2016,moshwan2017}. The values of the band gap and effective masses and the strength of electron-phonon coupling could change considerably near the topological phase transition, thus modifying the transport properties,  as demonstrated recently for PbSe
under external pressure~\cite{Chen2019}. A recent first principles study of thermoelectric transport in p-type SnTe argued that its almost linear bulk band dispersion could lead to a large enhancement of the Seebeck coefficient in nanostructures via energy filtering~\cite{Liu879}. However, the temperature induced changes of the electronic states in SnTe and their effect on the electronic lifetimes and transport have not been investigated, even though they could be more dramatic than in related NIs PbTe~\cite{cao2020} and SnSe~\cite{caruso2019}. 

Temperature-induced topological phase transitions in other material systems have been predicted from first principles~\cite{louie2016,Monserrat2016,peng2019,Hwan2020} and model Hamiltonians~\cite{garate2013,garate2014}. First principles approaches commonly use the standard Allen-Heine-Cardona (AHC) formalism~\cite{allen1976,allen1981,allen1983,allen1981}, in which the lattice thermal expansion and electron-phonon contributions to the band renormalization are treated independently. The AHC approach describes correctly the band gap changes with temperature in many materials with trivial topology~\cite{Giustino_diamond,antonius2014,ponce2015,prmaterialsPbTe}. However, this theory does not capture the influence of variations of electronic states with temperature on electron-phonon interaction, which may be important in materials with band inversion. Indeed, it has been shown recently that the renormalized electron energies computed using the AHC method in topological materials exhibit unphysical features e.g.~the valence and conduction bands remain parabolic even when the gap closes~\cite{Hwan2020}. Taking into account the thermal changes of electronic states is therefore necessary for an accurate description of electron-phonon interaction and the resulting band renormalization in topological insulators.

In this paper, we explicitly account for changes of the electronic states due to thermal expansion in calculating the temperature dependence of the band structure and electron-phonon interaction in SnTe. We show that temperature drives SnTe towards the phase transition from the TCI to NI phase, due to both thermal expansion and electron-phonon interaction. However, the topological phase transition does not occur before the melting temperature of SnTe ($\sim 1000$~K~\cite{tanaka2012}) is reached. We find that the band gap saturates with temperature, which cannot be captured using the AHC formalism. Moreover, we predict a remarkable non-monotonic temperature dependence of the lifetimes of the conduction states near the band edge as the gap decreases. Our findings open new avenues for accurate predictions and manipulation of bulk transport properties of topological materials.

\section{Methodology}\label{calculation}

We compute the ground state bulk electronic band structure of SnTe using density functional theory (DFT)~\cite{payne1992}, and then calculate the quasiparticle band structure renormalized by electron-electron and electron-phonon interaction, henceforth EEI and EPI, respectively. We consider the high-temperature rocksalt structure of SnTe, which gives rise to the topological crystalline phase~\cite{Hsieh2012}. We note that SnTe can undergo a ferroelectric structural phase transition below 100 K, depending on the doping concentration~\cite{dolling1966,brillson1974,iizumi1975,kobayaski1976,oneills2017}. Spin-orbit interaction is included in all our calculations.

\subsection{Ground state and GW electronic bands}

DFT ground state simulations are carried out with the {\sc Quantum ESPRESSO} suite~\cite{QE-2017, QE-2009}, using the Perdew-Burke-Ernzerhof (PBE) parametrization of the generalized gradient approximation (GGA)~\cite{Perdew1996} for the exchange and correlation functional. We use fully relativistic and fully nonlocal two-projector norm-conserving pseudopotentials from the {\sc PseudoDojo} data base~\cite{VanSetten2018}, generated with the the  Optimized  Norm-Conserving  Vanderbilt Pseudopotential ({\sc ONCVPSP}) code~\cite{Hamann}. All pseudopotentials include a full semicore shell. Wave functions are expanded in a basis of plane waves with a cutoff energy of 100 Ry. Brillouin zone sampling is carried out using a Monkhorst-Pack mesh \cite{Monkhorst1976} of $12\times12\times12$ reciprocal space $\one$-points. 

Electron-electron renormalization of the bulk band structure of SnTe is calculated using the G$_0$W$_0$ approximation~\cite{[][{, and references therein.}]Aryasetiawan_1998}. The G$_0$W$_0$ corrections are computed using the {\sc Yambo} code \cite{Sangalli_2019}, on a $16\times16\times16$ $\one$-point grid in the Brillouin zone. A cutoff energy of 30 Ry for the reciprocal lattice $\mathbf{G}$-vectors is used for the calculation of the exchange self-energy. The correlation part of the self-energy is obtained summing over 100 bands (out of which 46 bands are occupied), and screening is calculated within the plasmon-pole approximation using 120 bands.

\subsection{Temperature renormalization of electronic states}

To compute the effect of thermal expansion on the electronic structure of SnTe, we perform DFT electronic band calculations for the lattice parameters at different temperatures. We obtain the linear thermal expansion coefficient of SnTe by calculating its bulk modulus, the heat capacity and Gr\"uneisen parameters for the phonon modes in the entire Brillouin zone~\cite{pavone1994, prmaterialsPbTe, murphy2017}. Our computed thermal expansion coefficient agrees very well with the reported measurements between 0 K and 300 K~\cite{Smith_1976} (see Supplemental Material~\cite{supp}).
Using the experimental lattice constant at 300 K (6.327 \AA \cite{Smith_1976,bis1969}) and the calculated thermal expansion coefficient, we can obtain the lattice constant at any temperature~\cite{pavone1994, prmaterialsPbTe} (see Supplemental Material~\cite{supp}), as well as the corresponding electronic band structure.

We obtain the electronic band shifts due to electron-phonon interaction at a certain temperature using the DFT band structure that corresponds to the thermally expanded lattice at that temperature. As a result, the influence of thermal-expansion-induced band structure changes on electron-phonon interaction is included in our approach. We calculate the Fan-Migdal electron-phonon self-energy given as~\cite{cannuccia2013,antonius2015,Giustino2017}:   
\begin{eqnarray}\label{self-energy}
\Sigma_{n\one}^{FM} (\varepsilon_{n\one}, T) & = & \sum_{n' \two \lambda} \frac{\vert g_{nn'\one}^{\two \lambda}\vert^{2}}{N_\two}    \times\left[  \frac{n_{\two \lambda} (T) +1 -f_{n' \one + \two}(T)}{\varepsilon_{n\one} - \varepsilon_{n' \one + \two} - \omega_{\two \lambda} + i\delta }  \nonumber \right.  \\
& &  \left. + \frac{n_{\two \lambda} (T) + f_{n' \one + \two}(T)}{\varepsilon_{n\one} - \varepsilon_{n' \one + \two} + \omega_{\two \lambda} + i\delta} \right],\label{FM}
\end{eqnarray}
where $\varepsilon_{n\one}$ denotes the energy of the electronic state with the band index $n$ and crystal momentum $\one$, and $ \omega_{\two \lambda} $ stands for the phonon frequency with the branch index $\lambda$ and wavevector $\two$. $f_{n' \one + \two} (T)$ and $n_{\two \lambda} (T)$ are the Fermi-Dirac and Bose-Einstein distribution functions at temperature $T$, respectively. $N_{\two}$ is the total number of $\two$ vectors used for the integration, and $\delta$ is a broadening parameter. $g_{nn'\one}^{\two \lambda}$ is the first-order electron-phonon matrix element quantifying the probability amplitude for an electron in an initial state $n\one$ to be scattered to a final state $n\one+\two$ by emitting or absorbing a phonon $\two\lambda$~\cite{cannuccia2013,Giustino2017}:
\begin{equation}
g_{nn'\one}^{\two \lambda} = \langle u_{n'\one + \two} \vert\partial v^{KS}_{\two\lambda} \vert u_{n\one} \rangle_{\text{uc}}.\label{matrix_element}
\end{equation}
$\vert u_{n\one} \rangle$ denotes the lattice-periodic part of the wavefunction for the state $\vert {n\one} \rangle$, and the integral above is evaluated over the unit cell. $\partial v^{KS}_{\two \lambda}$ represents the variation of the Kohn-Sham potential due to the atomic displacements corresponding to the phonon mode $\two\lambda$~\cite{Giustino2017}:
\begin{equation}
\partial v^{KS}_{\two  \lambda} = \sqrt{\frac{\hbar}{2\omega_{\two  \lambda}}} \sum_{\kappa  \alpha} {\sqrt{\frac{1}{M_{\kappa}}}} e_{\kappa \alpha}^{
\lambda} (\two)\partial_{\kappa  \alpha, \two} v^{KS},
\end{equation}
where $\hbar$ is the reduced Planck constant, and $e_{\kappa  \alpha}^{\lambda} $ is the $\alpha$-th Cartesian component of the phonon eigenvector for an atom  $\kappa$ with mass $M_{\kappa}$. $\partial_{\kappa  \alpha, \two} v^{KS}$ is the lattice-periodic part of the perturbed Kohn-Sham potential expanded to first order in the atomic displacement~\cite{Giustino2017}.

The electron-phonon matrix elements $g_{nn'\one}^{\two \lambda}$ are calculated using density functional perturbation theory (DFPT) as implemented in {\sc Quantum ESPRESSO}~\cite{QE-2017, QE-2009}, using the GGA-PBE functional, 8$\times$8$\times$8 $\one$ and $\two$ grids and a cutoff energy of 100 Ry. We then use the real-space Wannier functions approach~\cite{Giustino2007,Giustino2017} and the {\sc EPW} code\cite{Ponce2016} to interpolate these matrix elements on finer 100$\times$100$\times$100 $\two$ grids for the electronic states along the W-L-$\Gamma$ path, employing 18 Wannier orbitals for interpolation. The electron-phonon self-energy in Eq.~\eqref{FM} is calculated non-self-consistently. Gaussian broadening is used to compute the imaginary part of electron-phonon self-energy (linewidth). The broadening parameter $\delta$ of $25$~meV gives converged values of real and imaginary parts of self-energy, and is used in all our calculations. To approximately account for the Debye-Waller shift of the real part of self-energy, the {\sc EPW} code enforces a sum rule to conserve the number of electrons: $\Sigma'_{n\one}(\omega,T) = \Sigma_{n\one}(\omega,T)-\Re\Sigma_{n\one}(\omega=\epsilon_F,T)$. $\Re\Sigma_{n\one}(\omega,T)$ is the real part of the Fan-Migdal self-energy and $\epsilon_F$ is the Fermi level. The sum rule thus requires that the total real part of the self-energy vanishes at the Fermi level, here taken to be in the middle of the gap.

We note that our computed transverse optical (TO) phonon frequencies decrease with increasing lattice volume due to thermal expansion, and become imaginary for the lattice constants above 550~K.  
Other phonon modes are not very sensitive to volume. Soft TO phonons are indicative of the ferroelectric phase transition in SnTe, and a similar volume dependence of its TO frequencies to that reported here has already been obtained in DFT calculations~\cite{ribeiro2018}. However, it has been experimentally observed that the TO mode frequency increases with temperature in the high-temperature phase~\cite{dolling1966,oneills2017},  due to strong anharmonicity that is not taken into account in our calculations~\cite{ribeiro2018}. To partially overcome this issue, we use the phonon energies and eigenvectors calculated for the experimental volume at 300~K to obtain the electron-phonon self-energy at all temperatures above 300 K. Such phonon dispersion is in very good agreement with inelastic neutron scattering~\cite{Cowley_1969, dolling1966} and inelastic X-ray scattering experiments~\cite{oneills2017} between 100 K and 300 K (see Supplemental Material~\cite{supp}).

\section{Results}\label{results}

\subsection{Ground state and G$_0$W$_0$ electronic band structure}

\begin{figure}
  \begin{center}
  \includegraphics[width=1.0\linewidth]{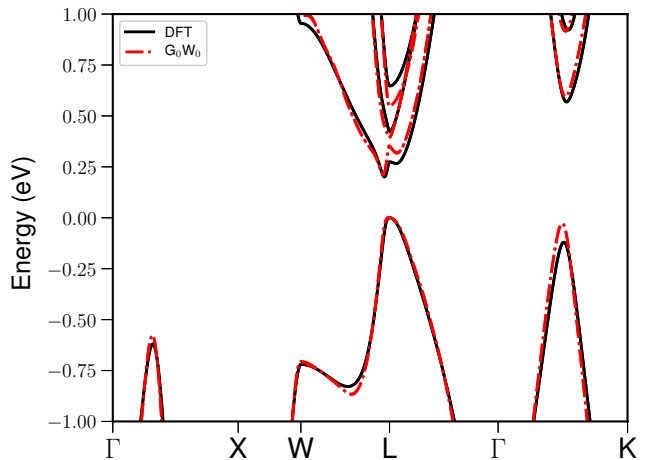}
  \caption[]{ Electronic band structure of SnTe obtained using density functional theory (DFT) within the generalized gradient approximation (solid black lines) and the many-body G$_{0}$W$_{0}$ approximation (dash-dotted red lines) for the lattice constant at 0~K (6.295 \AA). The energy of the valence band maximum at L is set to $0$ eV.}
  \label{Figure1}
  \end{center}
  \end{figure}

We first evaluate whether our DFT calculations give a reasonably accurate representation of the electronic band structure of SnTe. Fig.~\ref{Figure1} shows the comparison between the band structure obtained with DFT and the G$_{0}$W$_{0}$ approximation using the computed lattice constant at 0~K (6.295 \AA). The overall agreement between the DFT and G$_{0}$W$_{0}$ band structures is good.  Most importantly, the band ordering at L does not change when EEI is taken into account. The inclusion of electron-electron interaction in the G$_{0}$W$_{0}$ approximation leads to an increase of the fundamental gap from 0.20 to 0.216~eV, while the direct gap at L varies from 0.275 to 0.353~eV. These results
show that EEI favors the TCI phase. EEI does not change the band curvature much near the valence band maximum at L and the conduction band minimum near L. Both DFT and G$_{0}$W$_{0}$ calculations give band gaps and effective masses in reasonable agreement with experiments (see
Supplemental Material~\cite{supp}). Since the important features of the electronic band structure of SnTe are well captured using DFT, we will use it to calculate the temperature renormalization of the electronic band structure. 

\begin{figure}
\begin{center}
\includegraphics[width=0.48\linewidth]{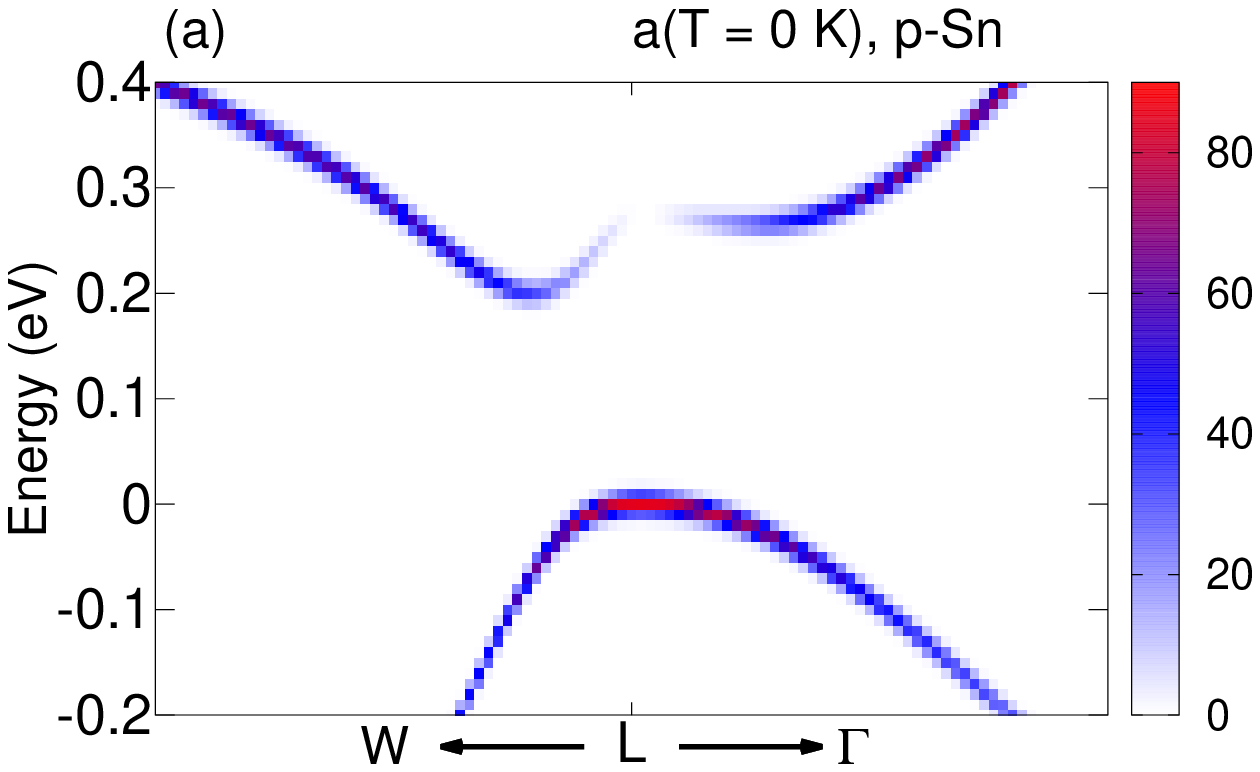} \hspace{1mm}
\includegraphics[width=0.48\linewidth]{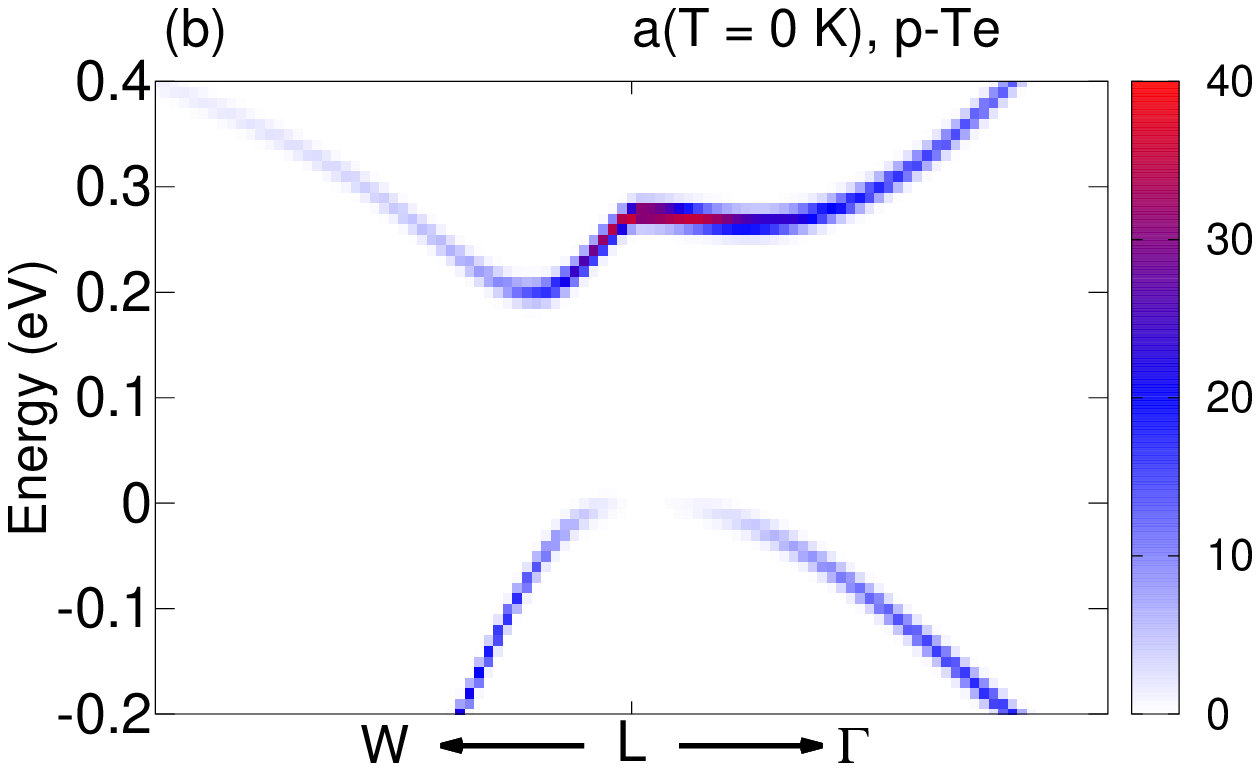}\\
\includegraphics[width=0.48\linewidth]{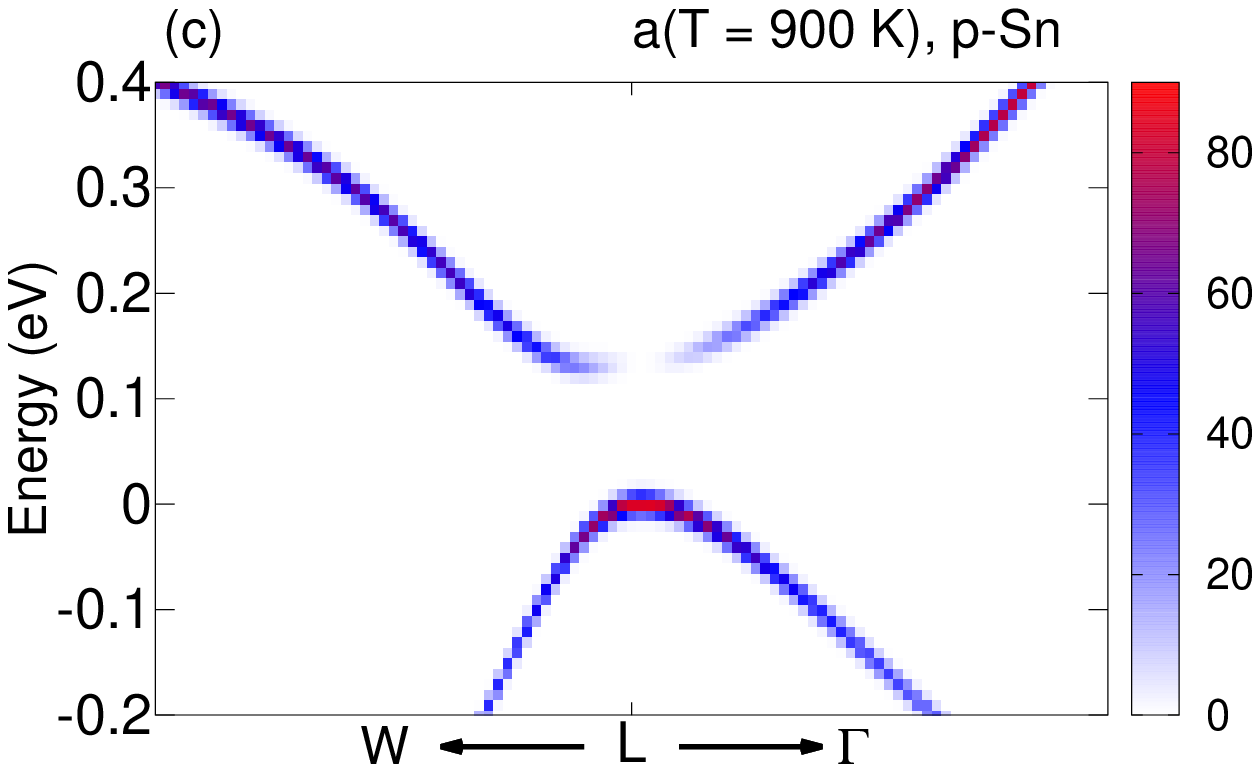}  
\hspace{1mm}
\includegraphics[width=0.48\linewidth]{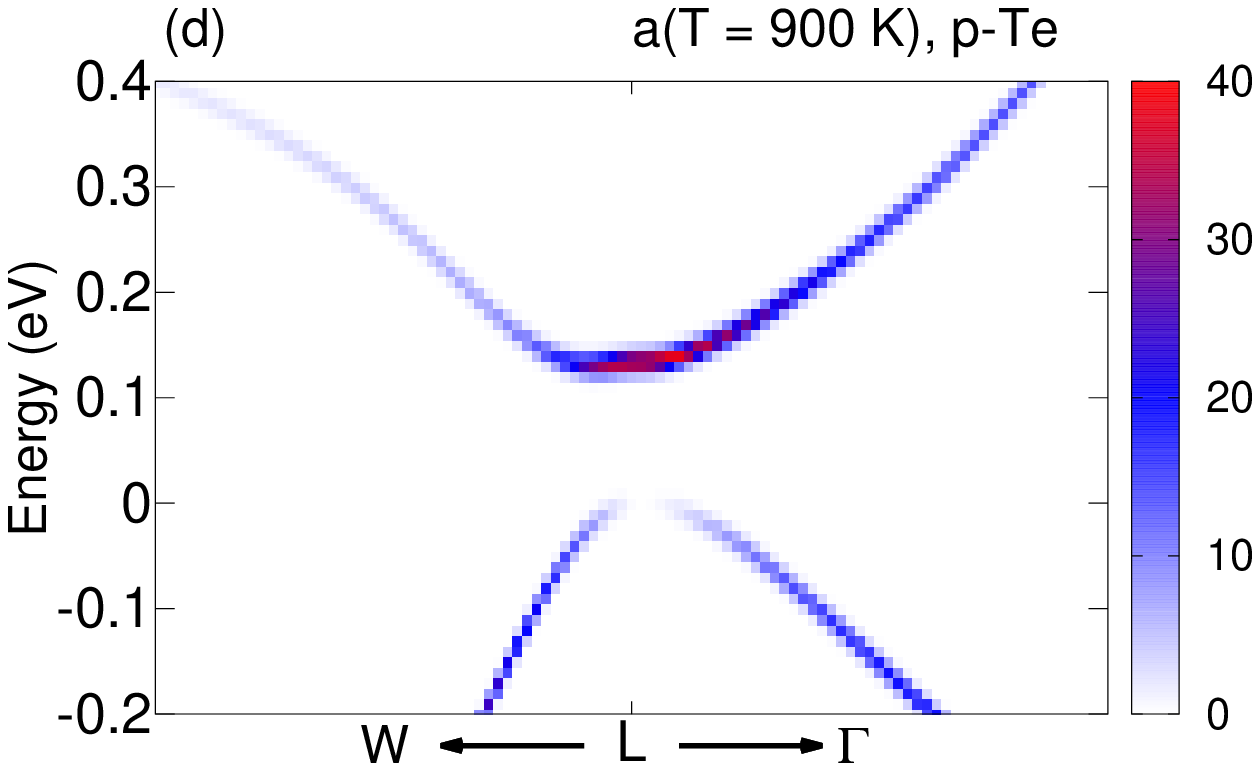}
\end{center}
\caption{Electronic band structure of SnTe calculated using density functional theory. The energies in (a) and (b) are for the lattice constant at T = 0 K and those in (c) and (d) are for T = 900 K. The color scale in (a) and (c) represents the projection onto the Sn $p$-orbitals and in (b) and (d) the projection onto the Te $p$-orbitals. The fractions of the L-W and L-$\Gamma$ lines shown are $1/4$ and $1/4$, respectively. The energy of the valence band maximum at L is set to $0$ eV.
}
\label{projection}
\end{figure}

To confirm that the electronic band structure of SnTe indeed exhibits band inversion, we show the projections of the DFT band states for the $0$~K lattice constant onto the Sn and Te p-orbitals in Figs.~\ref{projection}(a) and (b), respectively. Far enough from the L points, the conduction band states mostly have the character of the Sn $p$-orbitals, while their character changes to that of the Te $p$-orbitals near L. The valence band states show the opposite trend. As a result, the conduction band states close to L and the valence band states away from L have the character of the Te $p$-orbitals, while the valence band states near L and the conduction band states away from L have the character of the  Sn $p$-orbitals. Consequently, the ordering of the conduction and valence band states in the vicinity of the four L points is inverted, which is characteristic of the TCI phase. In SnTe, the states near L with the dominant character of the Te-$p$ orbitals are strongly repelled upward by the low-lying states with the character of the Sn-$s$ orbitals~\cite{zunger1997}. This effect pushes the Te-$p$ states above the Sn-$p$ states near L and causes the band inversion~\cite{barone2013}. At the L points, the valence and conduction band states correspond to the representations $L^{6-}$ and  $L^{6+}$, respectively~\cite{rabii1969}.

\subsection{Effect of thermal expansion on the electronic band structure}

\begin{figure}[h]
  \begin{center}
  \includegraphics[width=1.0\linewidth]{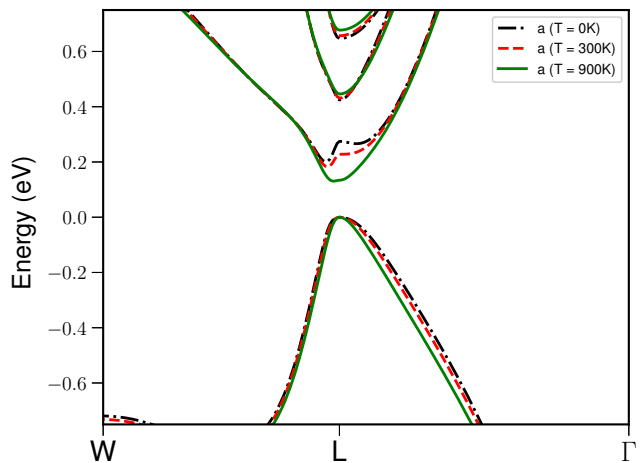}
  \caption[]{Electronic band structure of SnTe calculated using density functional theory for the lattice constants at 0 K, 300~K and 900 K. The energy of the valence band maximum at L is fixed at 0 eV.}
  \label{Figure3}
  \end{center}
  \end{figure}

The temperature dependence of the electronic states in SnTe arising from thermal expansion alone, without electron-phonon contributions, is illustrated in Fig.~\ref{Figure3}. We show the DFT electronic band structure computed for the lattice constants at 0~K, 300 K and 900~K. Thermal expansion decreases the direct gap in SnTe, and makes the valence band dispersion more linear and Dirac-like with increasing temperature. The conduction band dispersion is very non-parabolic at 0 K, with three extrema around L and the minimum located away from L (``Mexican-hat''-like). With thermal expansion, the conduction band dispersion becomes more parabolic i.e. it evolves towards the dispersion with a single minimum at L. Eventually, the conduction band becomes more linear for the lattice constants corresponding to temperatures larger than 900 K. These band structure features suggest that thermal expansion in SnTe tends to promote the normal insulating phase, without band inversion. This behavior is consistent with our previous results for the temperature dependence of the direct gap in PbTe~\cite{prmaterialsPbTe}. PbTe does not have band inversion and thermal expansion increases its direct band gap at L. Since the order of the valence and conduction bands in SnTe is inverted with respect to PbTe, the band gap of SnTe decreases with increasing thermal expansion.

Next we analyze in more detail how the band inversion near the L points of SnTe is affected by thermal expansion. Fig.~\ref{projection} shows the projections of the DFT band states onto the Sn and Te $p$-orbitals for the lattice constants at 0 K (top panels) and 900 K (bottom panels). All the plots are given in the same scales. The size of the region in $\one$-space in which the band inversion occurs decreases with thermal expansion. The reduction of the band inversion region is more prominent for the lattice constants above 900 K. These results confirm that thermal expansion drives SnTe closer to the topological phase transition. This effect is the consequence of the weaker repulsion between the Te-$p$ and Sn-$s$ states near L with the increasing bond length due to thermal expansion~\cite{zunger1997}, resulting in smaller direct band gaps close to L and larger direct band gaps away from L.

\begin{figure}
  \begin{center}
    \includegraphics[width=1.0\linewidth]{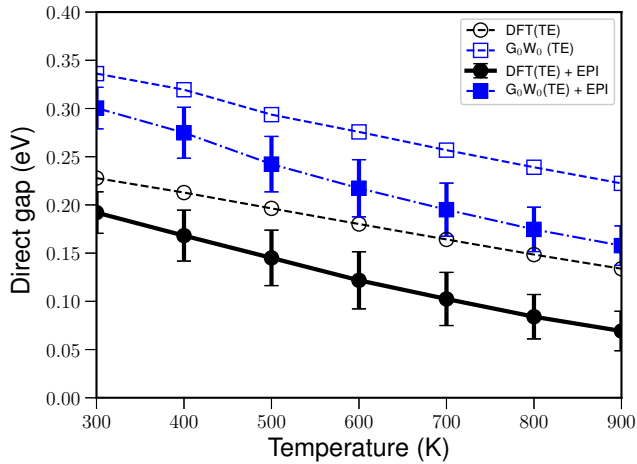}
    \caption[]{The direct band gap of SnTe at L as a function of temperature. Black and blue circles correspond to density functional theory (DFT) and G$_{0}$W$_{0}$ values, respectively. Open circles/squares illustrate DFT/G$_{0}$W$_{0}$ results including thermal expansion (TE) only, while full circles/squares represent DFT/G$_{0}$W$_{0}$ band gaps renormalized by both thermal expansion and electron-phonon interaction (EPI). The error bars are determined by the sum of the linewidths of the conduction and valence band states at L due to electron-phonon interaction.
}
  \label{directgap}
  \end{center}
  \end{figure}

We also compute the renormalization of the direct gap at L due to electron-electron interaction, while accounting for thermal expansion. We focus on the high temperature range (300 - 900 K) where SnTe is a TCI. The calculated DFT and G$_{0}$W$_{0}$ direct gap values for the lattice constants at different temperatures are given by open circles and squares, respectively, in Fig.~\ref{directgap}. Both DFT and G$_{0}$W$_{0}$ predict the band gap reduction with increasing lattice expansion (or temperature). The G$_{0}$W$_{0}$ quasiparticle gap roughly corresponds to a rigid shift of the DFT gap by $\sim 0.1$~eV for any lattice constant value, thus increasing the size of the gap and preserving the TCI phase compared to DFT.

  \begin{figure*}[hbt!]
  \begin{center}
   \includegraphics[width=0.30\linewidth]{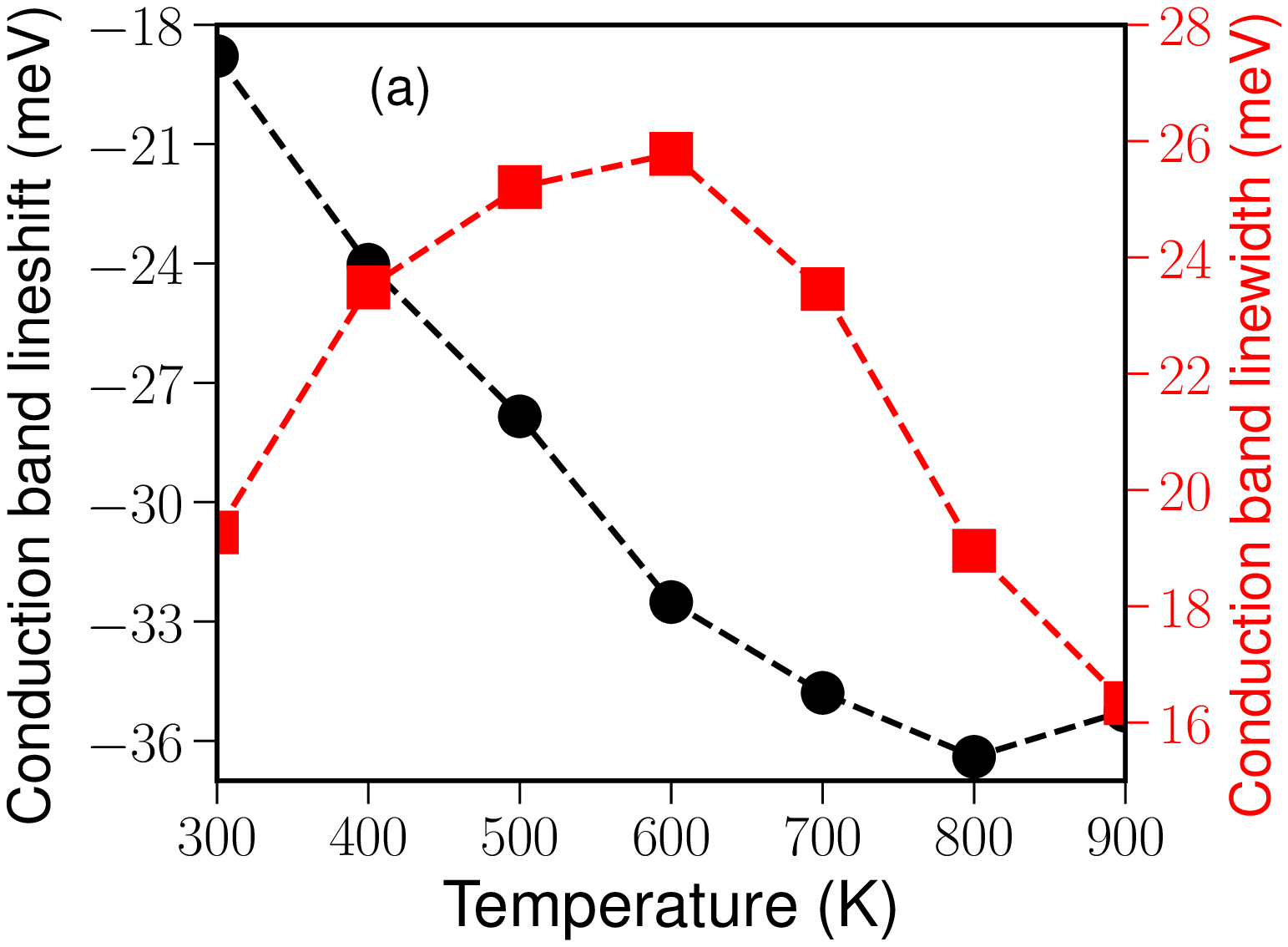}
    \includegraphics[width=0.295\linewidth]{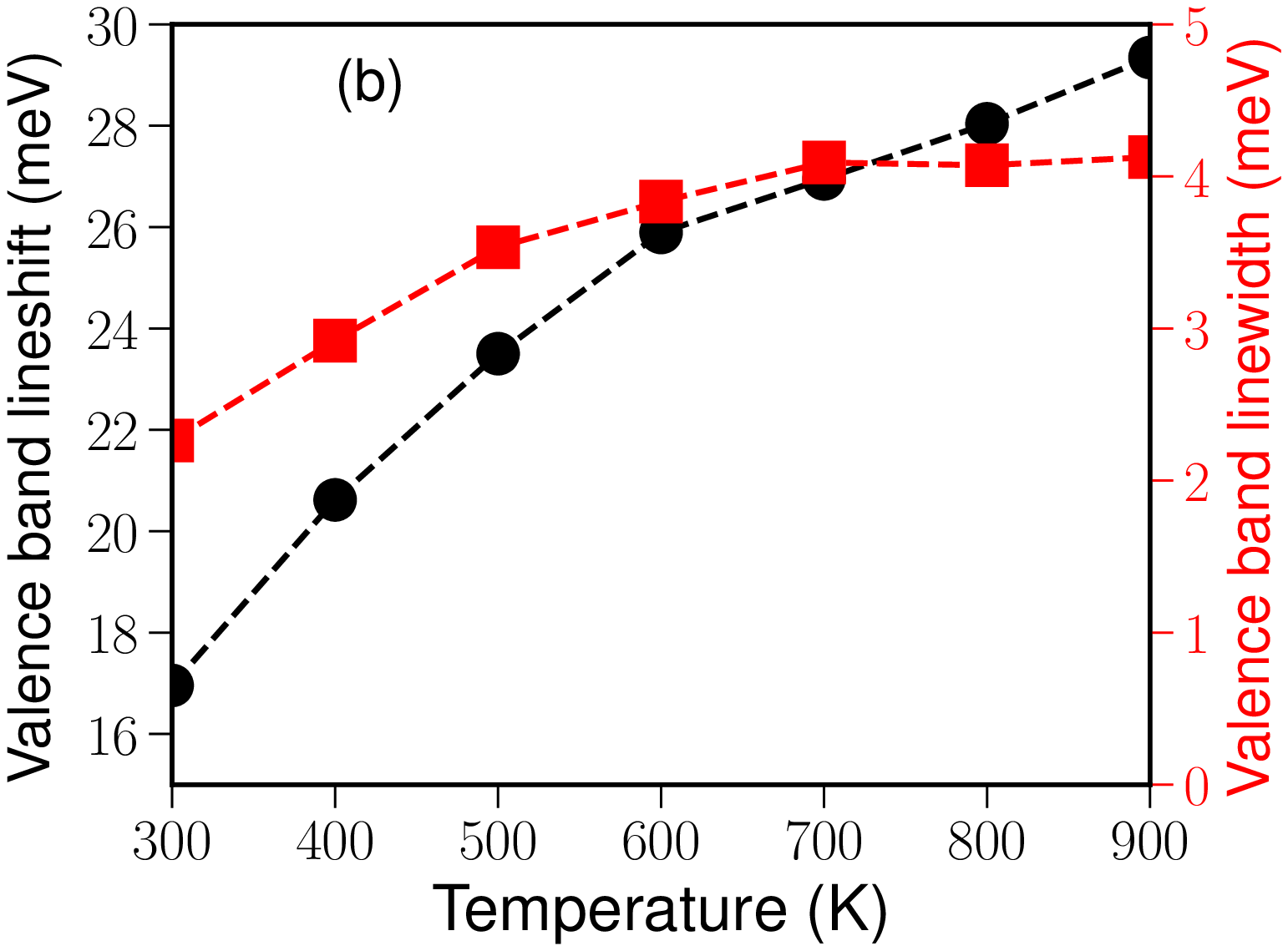}
  \includegraphics[width=0.30\linewidth]{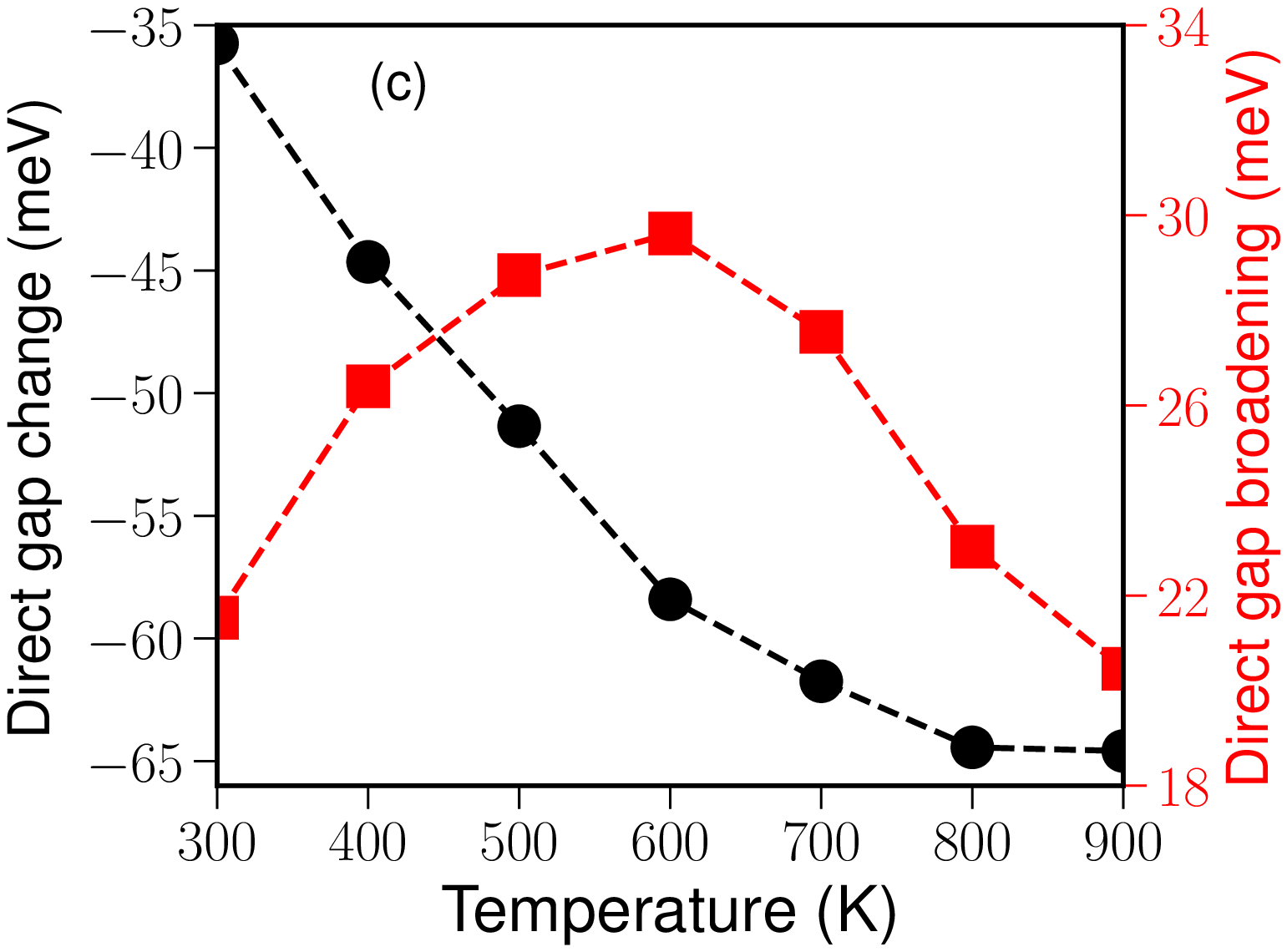}
  \caption[]{Lineshift (black circles) and linewidth (red squares) due to electron-phonon interaction of (a) the conduction band and (b) the valence band states at L in SnTe as a function of temperature. (c) Temperature variation (black circles) and broadening (red squares) of the direct gap at L. In all panels, black circles and red squares correspond to the vertical axis on the left and right, respectively. 
  }
  \label{directgap_braodening}
  \end{center}
  \end{figure*} 

\subsection{Variation of the electronic band structure with temperature}\label{total_gap_L}

Now we investigate how electron-phonon interaction renormalizes the direct band gap. Electron-phonon self-energy is calculated using the DFT band structure accounting for thermal expansion. Full circles and squares in Fig.~\ref{directgap} show how the DFT and G$_0$W$_0$ band gaps vary with temperature due to both thermal expansion and electron-phonon interaction. The error bars in Fig.~\ref{directgap} represent the uncertainty in the band gap values caused by EPI (which we will refer to as broadening), calculated as the sum of the linewidths (i.e.~the imaginary parts of the electron-phonon self-energy) of the conduction and valence band states at L at each temperature. We can see that electron-phonon interaction reduces the band gap and brings SnTe closer to the phase transition to a NI phase with increasing temperature. This trend is opposite in topologically trivial PbTe, where electron-phonon interaction increases the gap as temperature rises~\cite{prmaterialsPbTe}.

 \begin{figure}[hbt!]
  \begin{center}
  \includegraphics[width=1.0\linewidth]{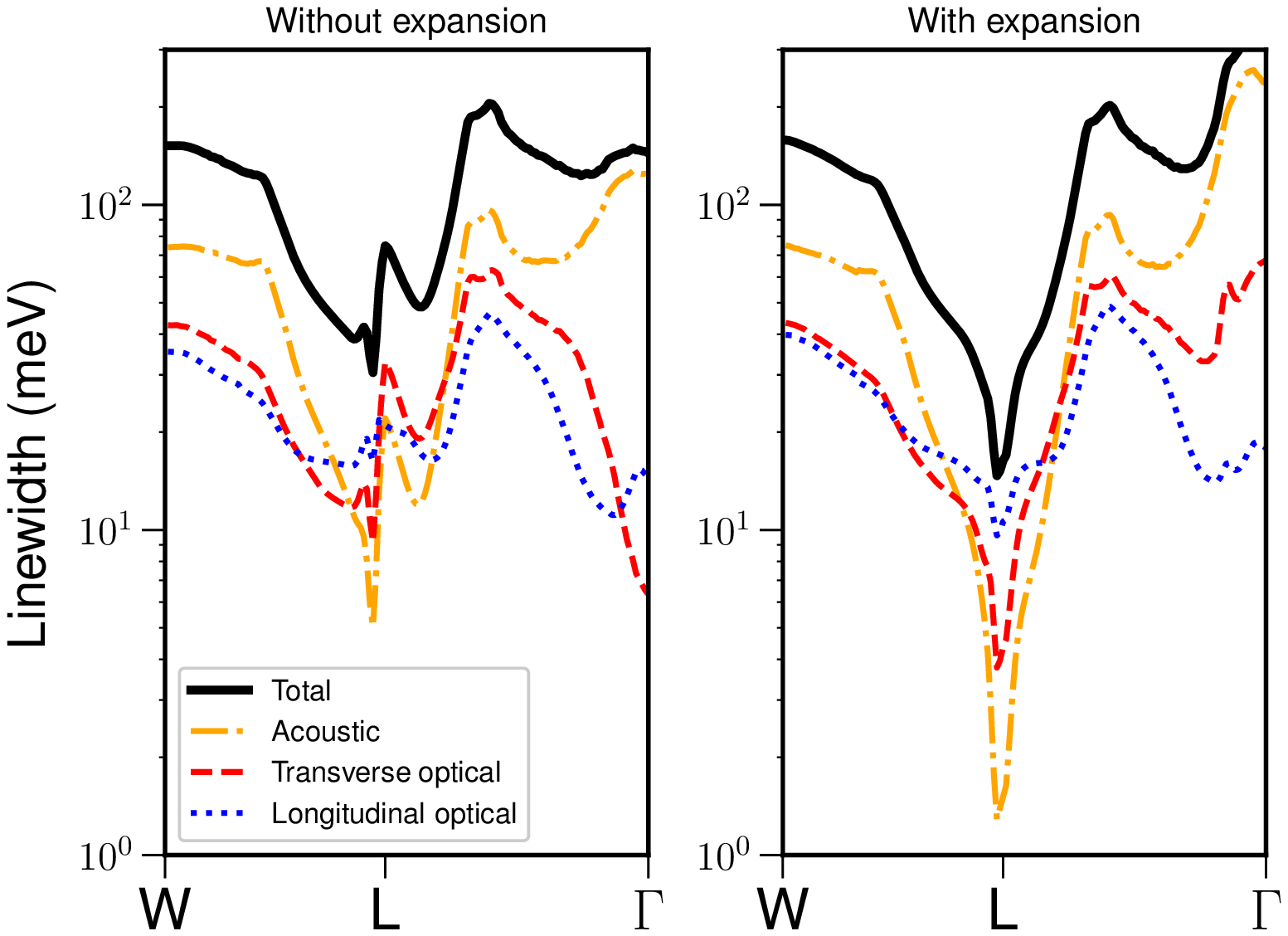}
  \includegraphics[width=1.0\linewidth]{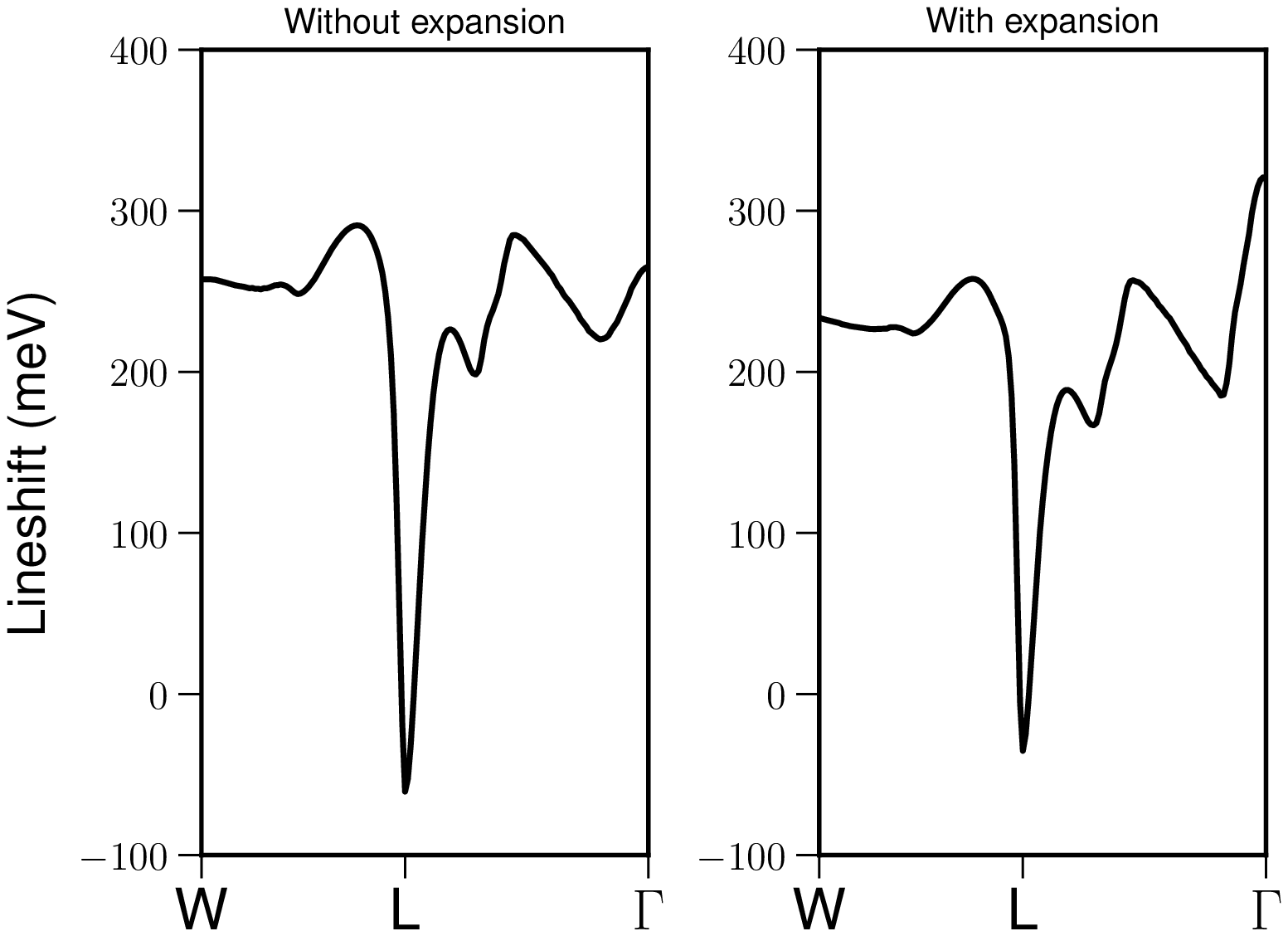}
  \caption[]{Effect of thermal expansion on the electron-phonon self-energy of the conduction band states in SnTe at 900 K. Top panels: linewidth resolved by phonon mode without (left panel) and with (right panel) thermal expansion included in the self-energy calculation. The contribution to the linewidth stemming from acoustic phonons is given by dash-dotted orange line, the transverse and longitudinal optical phonon contributions are represented by dashed red and dotted blue lines, respectively, while the total linewidth is given by solid black line. Bottom panels: total lineshift without (left panel) and with (right panel) thermal expansion.}
  \label{CB_selfnergy}
  \end{center}
  \end{figure}

      \begin{figure}[hbt!]
  \begin{center}
  \includegraphics[width=1.0\linewidth]{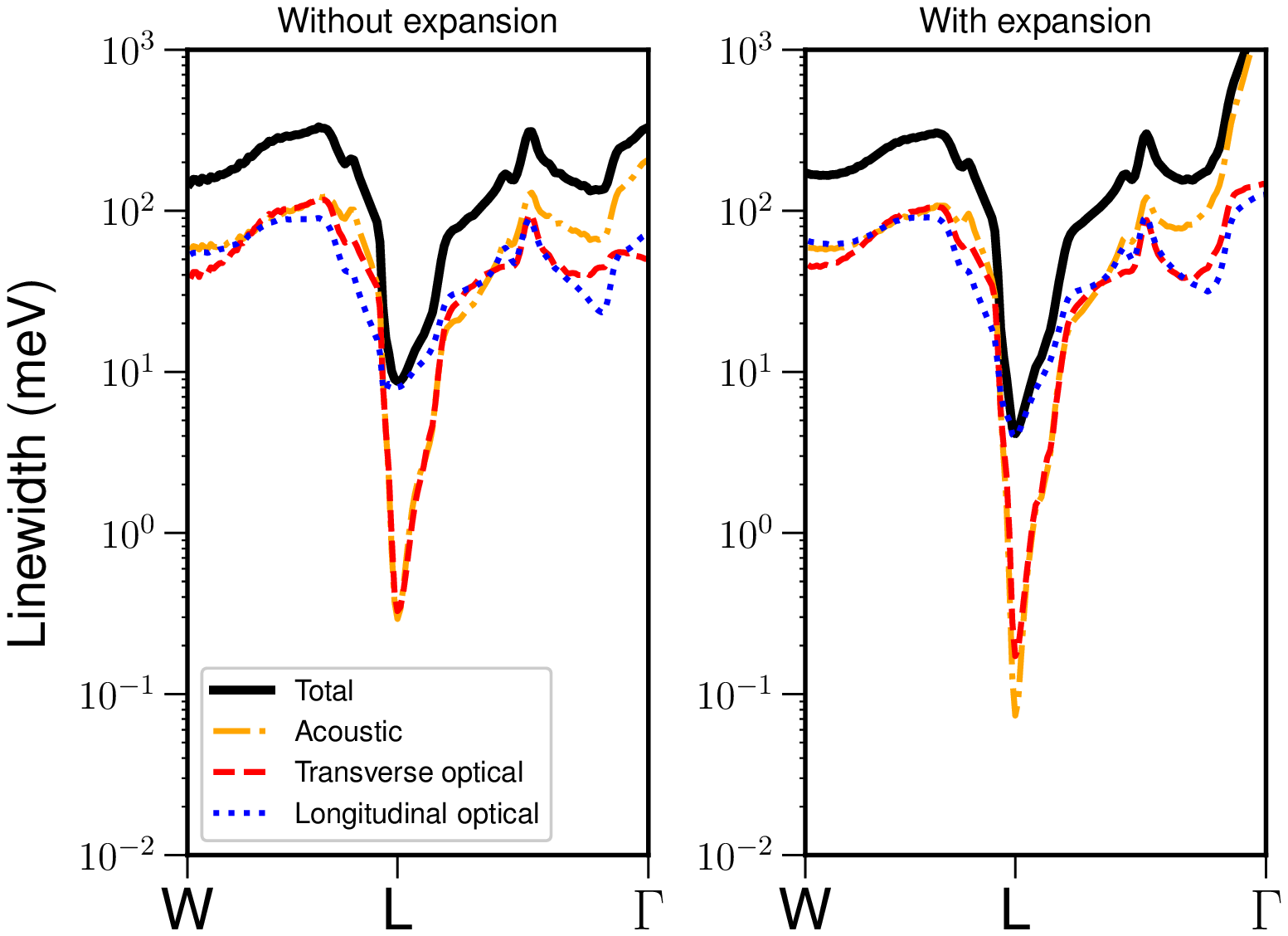}
  \includegraphics[width=1.0\linewidth]{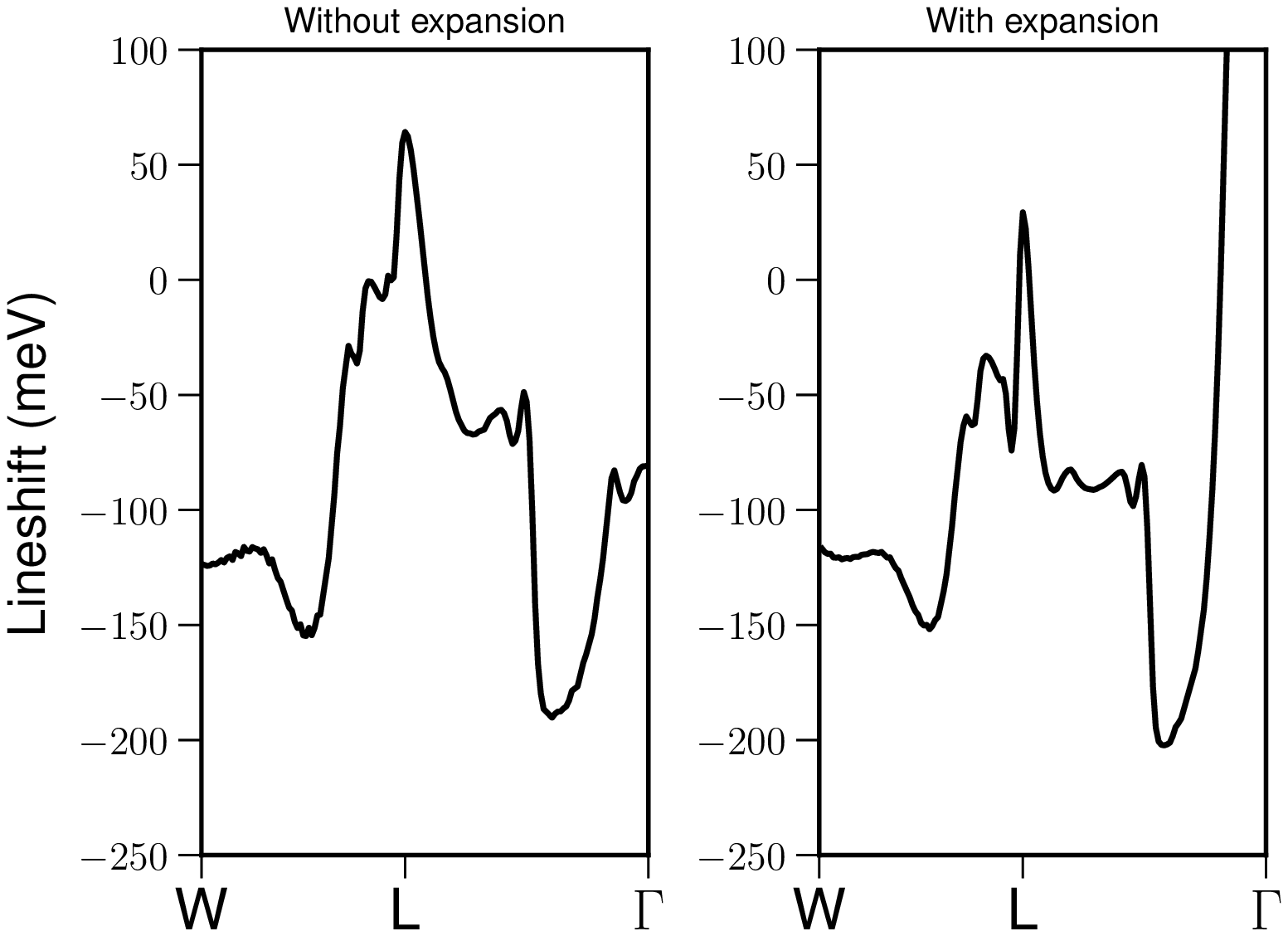}
  \caption[]{Effect of thermal expansion on the electron-phonon self-energy of the valence band states in SnTe at 900 K. All details are the same as in Fig. \ref{CB_selfnergy}.}
  \label{VB_selfnergy}
  \end{center}
  \end{figure}

The decrease of the direct band gap at L with temperature due to electron-phonon interaction is a direct consequence of the band inversion in SnTe. An electronic state is shifted downward (upward) due to coupling with another state of higher (lower) energy via phonons, see Eq.~\eqref{FM}. Since the band gap of SnTe is located near the four L points, the electron-phonon interaction is determined by the phonons whose wave vectors are close to the $\Gamma$ and X points. Due to the inversion center symmetry in SnTe, these phonons possess odd symmetry and couple electronic states with different parity~\cite{jiang2018}. Away from the L points, the conduction and valence band states mostly have the character of the Sn and Te $p$-orbitals, respectively, while their character is inverted near L. Therefore, the conduction band states near L are repelled downward due to the coupling with the conduction band states away from L, and upward due to the coupling with the valence band states close to L. The former effect is stronger as a result of the larger density of states and the smaller energy difference in the denominator of Eq.~\eqref{FM}, leading to a downward repulsion of the conduction band states near L. Similarly, the valence band states close to L are repelled upward, thus decreasing the direct band gap values near L with temperature due to EPI.
      
Our results show that both thermal expansion and electron-phonon interaction favor the temperature driven phase transition from the TCI to NI phase, in contrast to electron-electron interaction. All three effects give comparable contributions to the direct gap renormalization, see Fig.~\ref{directgap}, which result in the the negative temperature coefficient of the direct gap in SnTe. However, we predict that the band gap does not close up to 900~K. Since the melting temperature of SnTe is $\sim$ 1063 K~\cite{tanaka2012}, our calculations indicate that SnTe does not undergo the topological phase transition below the melting temperature. 
  
To understand better the contribution of electron-phonon interaction to the direct gap changes with temperature, we plot the temperature dependence of the lineshifts (the real part of electron-phonon self-energy) and linewidths of the valence and conduction band states at L (VBL and CBL, respectively), see Fig.~\ref{directgap_braodening}. Fig.~\ref{directgap_braodening}(c) shows the direct gap variation and broadening as a function of temperature. The temperature dependence of the CBL and VBL lineshifts and the band gap renormalization departs from linear, with the gap saturation at high temperatures. The variations of the CBL linewidth are more remarkable, increasing at low temperatures and decreasing at higher temperatures. The VBL linewidth rises at low temperatures and saturates at higher temperatures. The linewidth of VBL is much smaller than that of CBL due to the much smaller density of the valence states near L compared to the conduction states. This results in the similar trends and values for the gap broadening as those for the CBL linewidth. All these non-linear temperature trends predicted for the TCI phase of SnTe are not very typical for normal insulators without any changes in band topology~\cite{Giustino_diamond,antonius2014,ponce2015,prmaterialsPbTe}.
 

To elucidate the origin of the non-linear variations of the linewidths and lineshifts with temperature, we compute the electron-phonon self-energy along the W-L-$\Gamma$ path. The effect of thermal lattice expansion on the self-energy for the conduction and valence band states in SnTe at $900$ K is illustrated in Figs.~\ref{CB_selfnergy} and \ref{VB_selfnergy}, respectively. In both figures, we show the linewidths and lineshifts calculated for the lattice constants at 0~K and 900~K, while the temperature of phonon and electronic populations is set to 900 K for both lattice constant values. This is done to separate the band structure effects on the electron-phonon self-energy from the effect of phonon occupation changes with temperature (the effect of electron occupations is relatively small compared to that of phonons, see Eq.~\eqref{self-energy}). The linewidths of the inverted states are decreased when thermal expansion is included in the self-energy calculation (see solid black lines in top panels of Figs.~\ref{CB_selfnergy} and \ref{VB_selfnergy}). This reduction is much more pronounced for the conduction band states.

Our analysis reveals that the electron-phonon linewidth decrease with thermal expansion shown in Figs.~\ref{CB_selfnergy} and \ref{VB_selfnergy} stems from the reduced density of states near L and thus the reduced phase space for electron-phonon scattering. With thermal expansion, the valence band states become more linear and the effective masses defined in the energy range near the band edge at L decrease (see Supplemental Material~\cite{supp}). The conduction band states near L become less ``Mexican-hat''-like due to thermal expansion, exhibiting much larger decrease of the overall density of states in the energy range relevant for scattering compared to the valence bands. Therefore, the reduction of the linewidths close to L induced by thermal expansion is much more remarkable for the conduction band states compared to the valence band states. In contrast, the electron-phonon matrix elements of the conduction and valence states near L change very little with thermal expansion up to 900 K.

On the other hand, the linewidths of the conduction and valence band states change linearly with temperature if thermal expansion effects on electron-phonon interaction are neglected in our calculations. This is a consequence of the linear dependence of phonon occupations on temperature above the Debye temperature ($\sim$200 K for SnTe). As a result, the competition between the phonon populations and the thermal expansion effects on the electronic states determines the temperature dependence of the linewidths and lineshifts illustrated in Fig.~\ref{directgap_braodening}. At sufficiently high temperatures, the effect of the decrease in the density of states caused by thermal expansion on the self-energy of the conduction states near L becomes stronger than the effect of phonon occupations, leading to the non-monotonic temperature dependence of the linewidths and lineshifts (Fig.~\ref{directgap_braodening}(a)). The influence of the less dramatic thermal-expansion-induced reductions in the density of the valence states close to L on their self-energy becomes only comparable to the phonon population effects at certain temperatures, resulting in the non-linear variations of the self-energy with temperature (Fig.~\ref{directgap_braodening}(b)).

The lineshifts of the conduction and valence bands (bottom panels in Figs.~\ref{CB_selfnergy} and \ref{VB_selfnergy}) show two regions with different temperature behavior. The conduction band states have a negative lineshift in the vicinity of the L points where the band inversion occurs, while the lineshift becomes positive away from L. The opposite effect is found for the valence band states. Such momentum dependence of the lineshifts means that the direct gap values near L decrease, while further away from L the direct band increases. The latter effect occurs because the conduction band states away from L are repelled upward due to the coupling with the conduction band states near L and the valence band states away from L. Similarly, the valence band states away from L are repelled downward, resulting in the increase of the direct band gap away from L with temperature due to electron-phonon interaction. The change of the lineshift sign in the vicinity of L shows
that electron-phonon interaction also makes the electronic bands more linear at higher temperatures (in addition to thermal expansion shown in Fig.~\ref{Figure3}). The absolute lineshift values close to L are decreased if the effect of thermal expansion on the electronic states is included in the calculation, similarly to the linewidths. However, the linewidth reductions are more striking due to the energy conservation condition. 

\begin{figure}[hbt!]
\begin{center}
\includegraphics[width=1.0\linewidth]{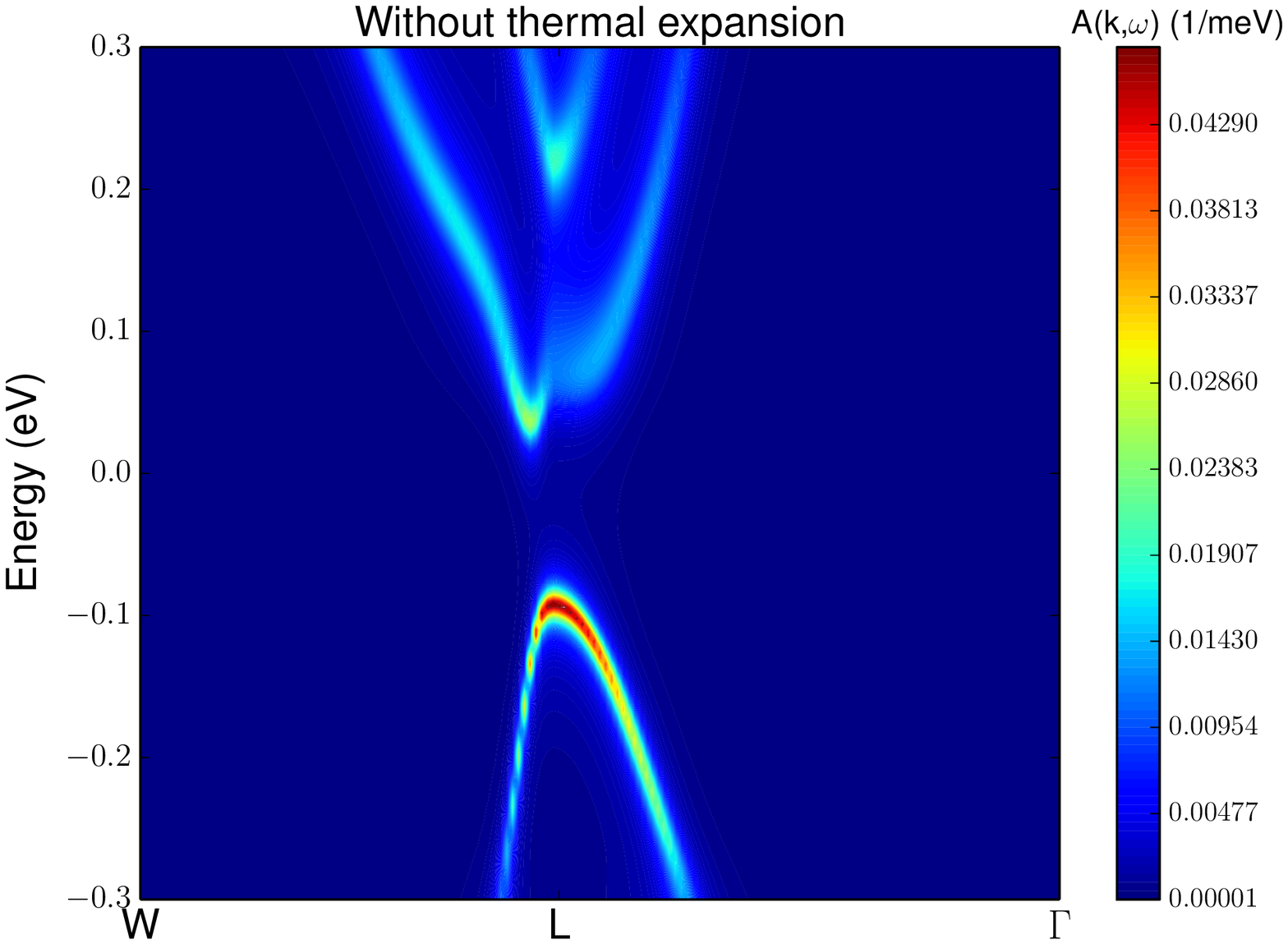}
\hspace{10mm}\includegraphics[width=1.0\linewidth]{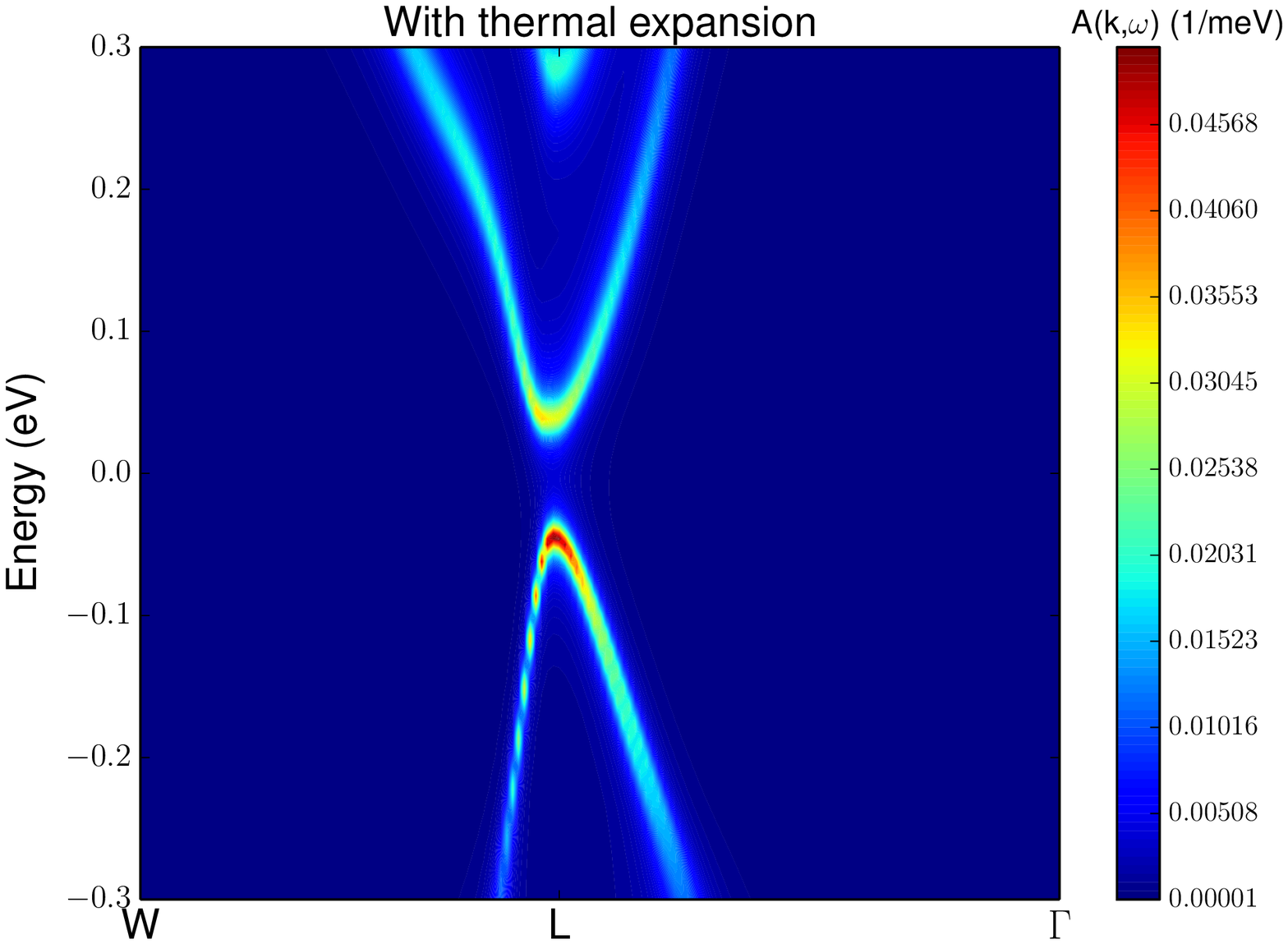}  \\
\end{center}
\caption{Effect of thermal expansion on the single particle spectral function $A(\one,\omega)$ of SnTe at 900 K. The calculation without thermal expansion (i.e.~using the lattice constant at 0~K) is given in the top panel, while the calculation with thermal expansion (i.e.~using the lattice constant at 900~K) is shown in the bottom panel.}\label{spectralfunction}
\end{figure}

We also show the contributions to the conduction and valence band linewidths resolved by phonon modes in top panels in Figs.~\ref{CB_selfnergy} and \ref{VB_selfnergy}, respectively. The relative contribution of different modes to the valence band linewidth is almost unaffected by thermal expansion. However, this is not the case for the conduction band, whose total linewidth also changes dramatically with thermal expansion. When the effect of thermal expansion on electron-phonon interaction is accounted for, longitudinal optical (LO) phonon scattering is the strongest scattering mechanism for both valence and conduction band states near L. This result is similar to our previous findings in n-type PbTe~\cite{cao2020,jiang2018}, where LO phonon scattering dominates electronic transport even at high doping concentrations and temperatures.

\subsection{Spectral function}\label{total_gap_sigma}

To get a complete picture of the electronic spectrum changes due thermal expansion and electron-phonon coupling, we also compute the single particle spectral function $A(\one,\omega)$ as:
\begin{equation}
    A_{n\one}(\omega,T) = \frac{1}{\pi}\frac{\lvert \Im\Sigma_{n\one}(\omega,T)\rvert}{\lvert\omega-\varepsilon_{n\one}-\Re\Sigma_{n\one}(\omega,T)\rvert^{2} + \lvert\Im\Sigma_{n\one}(\omega,T)\rvert^{2}}\label{sf}
\end{equation}
where $\hbar\omega$ is the binding energy, and $\Im\Sigma_{n\one}(\omega,T)$ is the imaginary part of the electron-phonon self-energy. The real and imaginary parts of the self-energy used to calculate the spectral function are obtained using Eq.~\eqref{FM} with the broadening parameter $\delta$ of $25$ meV.

Figure~\ref{spectralfunction} shows the spectral function calculated with the lattice constants at 0~K and 900~K, and the temperature of phonon populations of 900 K for both lattice constants. As already seen in Fig.~\ref{Figure3}, the band dispersion evolves towards more linear dispersion with thermal expansion. Electron-phonon interaction further contributes to this trend, see bottom panels in Figs.~\ref{CB_selfnergy} and \ref{VB_selfnergy}.  Moreover, thermal expansion reduces the linewidth of the conduction and valence band states around the L points, as observed in top panels in Figs.~\ref{CB_selfnergy} and \ref{VB_selfnergy}. All these features are reflected in the spectral functions given in Fig.~\ref{spectralfunction} and should lead to unusual temperature dependence of the electronic and thermoelectric transport properties of SnTe and other TCIs. 

\section{Discussion}\label{discussion}

Throughout this paper, we emphasize that it is essential to take into account the influence of thermal expansion on electronic states to obtain a more realistic prediction of the temperature behavior of electronic lineshifts and linewidths in topological materials. However, we neglect the effect of electron-electron induced band renormalization when calculating electron-phonon self-energy. We also ignore the fact that the electron-electron and electron-phonon self-energy should ideally be computed self-consistently. The self-consistent calculation of electron-electron effects would have been important if the mixing of the valence and conduction band states near the gap was substantially different in the G$_0$W$_0$ and DFT simulations, which is not the case here~\cite{aguado2020}. Furthermore, the effect of electron-electron band renormalization on electron-phonon self-energy should be mostly cancelled by the self-consistent procedure to calculate electron-phonon self-energy, because EEI and EPI induced shifts of the band gap are of the opposite sign and roughly cancel each other (see Fig.~\ref{directgap}). Nevertheless, it would be interesting to evaluate these effects in future work. In any case, our results indicate the importance of accounting for the temperature dependence of electronic structure when calculating electron-phonon self-energy in topological insulators.

\section{Summary and conclusions}\label{conclusions}

The temperature renormalization of the bulk electronic structure of SnTe in the topological crystalline phase has been investigated from first principles, including the effect of thermal expansion on the electron-phonon self-energy. Thermal expansion and electron-phonon interaction tend to decrease the band gap and bring SnTe closer to the phase transition to a normal insulating phase, in contrast to electron-electron interaction. The band gap shows a marked non-linear dependence on temperature as SnTe approaches the topological phase transition, which is accompanied by non-monotonic variations with temperature of the linewidths of the conduction band states near the band edge. These effects should have interesting consequences on the transport properties of SnTe. Our results show that electron-phonon interaction in topological materials is substantially affected by the thermal changes of their bulk electronic states. Our predictions could be tested using angle-resolved photoemission spectroscopy experiments in SnTe samples with small Sn vacancy concentrations\cite{tanaka2012} and other topological insulators.

\section{Acknowledgements}
 We thank G. Jeffrey Snyder and Felipe Murphy-Armando for insightful discussions.
 T.N.T., P.C. and M.G. thank Andrew Horsfield for a long-standing discussion on the 
perturbation of electronic states by phonons.
 This work is supported by Science Foundation Ireland and the Department for the Economy
 Northern Ireland Investigators Programme under grant number 15/IA/3160. I.S. also acknowledges
 support from  Science  Foundation Ireland and the European Regional Development Fund 
under grant number 13/RC/2077. We are grateful for the use of computational facilities 
at the Irish Centre for High-End Computing (ICHEC) and the UK national high performance computing service.

\bibliography{references} 

\end{document}